\documentclass[5p,12pt]{elsarticle}

\usepackage{graphicx}

\usepackage{amssymb}
\usepackage{siunitx}
\usepackage[official]{eurosym}

\usepackage{upgreek}

\journal{Advances in Space Research}

\graphicspath{{./figures/}}

\begin{document}

\begin{frontmatter}


\title{First successful satellite laser ranging with a fibre-based transmitter}

\author[]{D.~Hampf\corref{cor1}}
\ead{daniel.hampf@dlr.de}
\author[]{F.~Sproll\corref{cor1}}
\ead{fabian.sproll@dlr.de}
\author[]{P.~Wagner}
\author[]{L.~Humbert}
\author[]{T.~Hasenohr}
\author[]{W.~Riede}

\cortext[cor1]{Corresponding author}
\address{German Aerospace Center, Institute of Technical Physics, Pfaffenwaldring 38-40, 70569 Stuttgart}

\begin{abstract}
Satellite Laser Ranging (SLR) is an established technology used for geodesy, fundamental science and precise orbit determination. This paper reports on the first successful SLR measurement from the German Aerospace Center research observatory in Stuttgart. While many SLR stations are in operation, the experiment described here is unique in several ways: The modular system has been assembled completely from commercial off-the-shelf components, which increases flexibility and significantly reduces hardware costs. To our knowledge it has been the first time that an SLR measurement has been conducted using an optical fibre rather than a coud\'{e} path to direct the light from the laser source onto the telescope. The transmitter operates at an output power of about 75 mW and a repetition rate of 3 kHz, and at a wavelength of 1064 nm. Due to its rather small diameter of only 80 $\upmu$m, the receiver detector features a low noise rate of less than 2 kHz and can be operated without gating in many cases. With this set-up, clear return signals have been received from several orbital objects equipped with retroreflectors. In its current configuration, the system does not yet achieve the same performance as other SLR systems in terms of precision, maximum distance and the capability of daylight ranging; however, plans to overcome these limitations are outlined.

\end{abstract}


\end{frontmatter}

\parindent=0.5 cm

\section{Introduction}

Laser ranging to objects in space is of great importance in many different fields. By measuring the distance to dedicated satellites various scientific phenomena such as tectonic plate drifts, crustal deformation, the Earth's gravity field or ocean tides can be investigated \citep{DEG94}. Furthermore, precise orbit determination of cooperative as well as non-cooperative objects can support evasive manoeuvres to avoid collisions with space debris \citep{Bennett20131876}. Besides, re-entry predictions and mission preparation can benefit from satellite laser ranging (SLR) \citep{FLO15}. Together with VLBI (Very Long Baseline Interferometry) it is one of the core technologies for the GGOS (Global Geodetic Observing System) \citep{Rummel2005357}.

This variety of different SLR tasks requires a complete network of SLR stations. Whereas the coverage of stations on the northern hemisphere is sufficient at the moment, only a few stations are available on the southern hemisphere. Furthermore more stations will be required in the future due to the increasing number of satellites equipped with retroreflectors; also on the northern hemisphere. Especially for the rapidly increasing nano-satellite population, low cost and weight retroreflectors are attractive for orbit determination \citep{KIR13,BUC15,YOU15}. 

Since the first successful reported SLR experiment in 1964 \citep{XU10} the technological progress is tremendous. Commercial InGaAs based single photon detectors with low dark noise in the kHz regime and diameters in the range of \SI{100}{\micro\meter} are available \citep{HAD09,ROC09}. This enables detection of infrared photons up to \SI{1550}{\nano\meter} with efficiencies on the order of \SI{30}{\percent}. Thus the benefits of the fundamental Nd:YAG wavelength compared to its first harmonic for SLR can be exploited \citep{VOE13}. Furthermore the detectors can be operated in a non-gated mode due to their low dark count rates, at least during night time.

Additionally laser systems with \si{\kilo\hertz} repetition rate, picosecond pulse duration and several hundred micro joule pulse energy have been developed. Initially suggested by \cite{DEG94} first independent demonstrations of \si{\kilo\hertz} SLR systems were conducted in 2004  by an SLR2000 prototype and the Graz SLR station \citep{MCG04,KIR04}. Higher repetition rates allow lower pulse energies while still obtaining the same number of returns per time interval as in conventional designs \citep{DEG94,DEG94_2}. Theoretical considerations and experimental verifications show that pulse energies in the range of several ten micro joules are sufficient for SLR \citep{Leif2015, DEG94, MCG04, KIR15}; a big step towards new innovative SLR approaches. 

Novel laser transmitter designs can be realized without using a coud\'{e} path. Such a coud\'{e} path is the common solution in the SLR community for guiding the laser pulses from the source to the transmitter telescope on an astronomical mount. While enabling high laser pulse energies, the coud\'{e} path has several disadvantages. An expensive design as well as a crucial alignment of the mirrors are two main drawbacks. Smaller laser systems in the micro joule regime can be mounted directly onto the astronomical mount as proposed by \cite{DEG94} and realized for example by \cite{KIR15}. However, in this configuration the laser is exposed to the environment and various weather conditions. Furthermore the orientation of a laser attached directly to the mount changes continuously. To avoid these issues we prefer a fibre based laser transmitter. Laser pulses from a source in a controlled environment are guided via an optical fibre onto the mount. This enables an easy upgrade of existing small telescopes to SLR stations without any weight or size limitations of the laser system. Even light from several different laser sources for different tasks (e.g. LIDAR, optical communication) can be coupled into the fibre. Therefore a very modular and versatile system using only one telescope is possible with the fibre based approach.

In this paper we present the combination of an innovative multi mode fibre (MMF) based laser transmitter design with a low noise, non-gated infrared detection technology. We describe the system set-up (section \ref{hardware}), first experimental results (section \ref{results}) and discuss the potential of this technology for high-precision SLR (section \ref{fibre_based_SLR}).

\section{Hardware}
\label{hardware}

This section describes in detail the innovative set-up of the system, with the goal of making the replication of our system feasible for other parties. All components are readily available for purchase from various suppliers, with two exceptions: The beam splitter unit in the receiver channel has been designed and built by our institute, and the single photon detector has been discontinued by the manufacturer\footnote{However, commercial alternatives exist also for these two components, see section \ref{sec:receiver}.}.

With its small footprint, the complete system including laser and IT infrastructure, fits into a \SI{3.6}{m} clamshell dome. Total hardware costs are estimated to about 175 k\euro{} in 2016 prices. The modular approach used here is believed to be especially useful for the extension of existing, passive-optical systems. 

\subsection{Tracking}
\label{tracking}

\begin{figure}[tb]
   \centering
   \includegraphics[width=1\columnwidth]{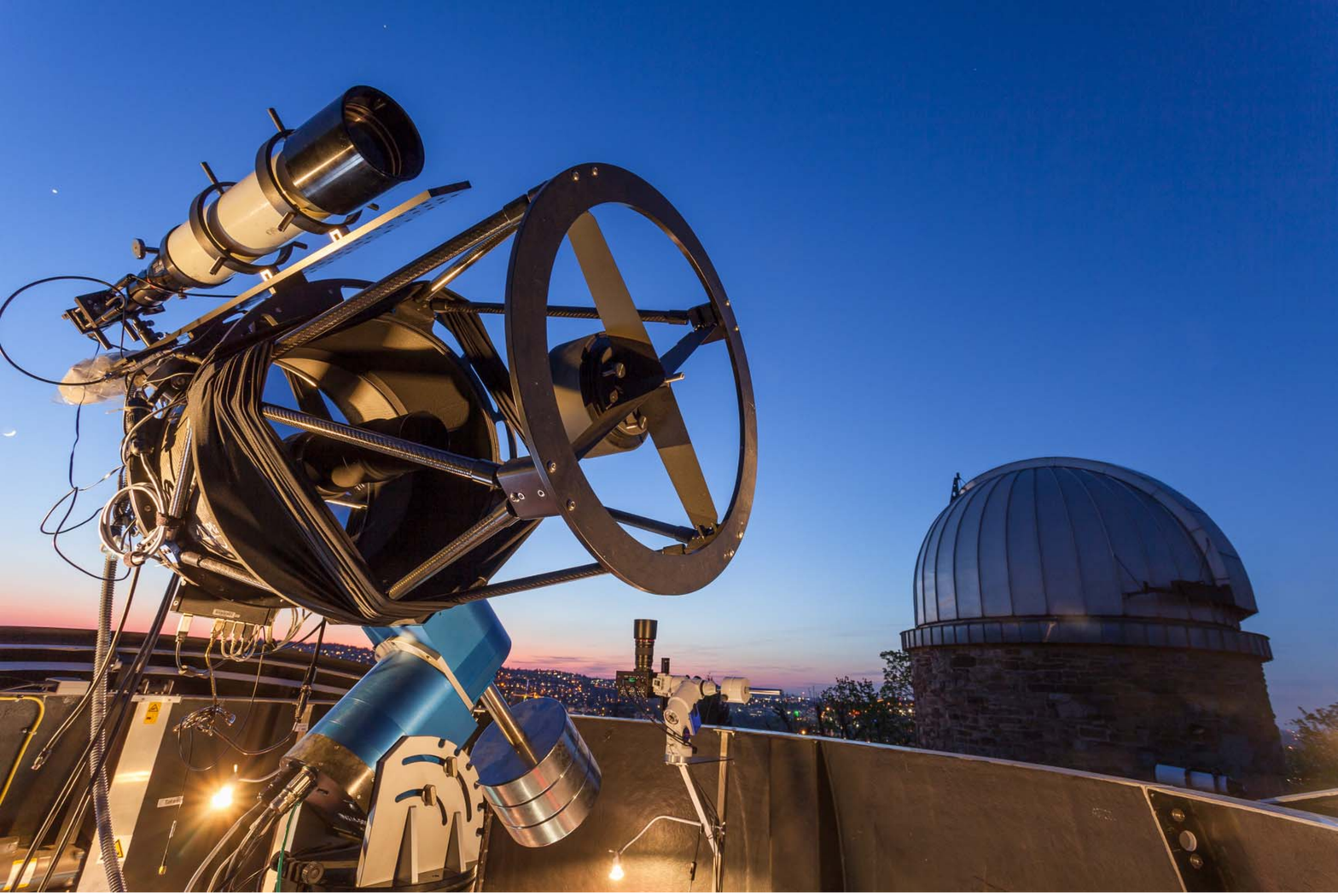}
   \caption{Telescope assembly used for laser ranging: The \SI{43}{\centi\meter} mirror telescope is used for tracking and as receiver for the laser photons, while the \SI{10}{\centi\meter} refractor telescope transmits laser pulses.}
   \label{telescope_image}
\end{figure}

Figure \ref{telescope_image} shows the tracking system consisting of a \SI{43}{\centi\meter} corrected Dall-Kirkham telescope\footnote{Planewave CDK 17, \url{http://planewave.com}}, an sCMOS camera\footnote{Andor Zyla 5.5 USB 3.0, \url{www.andor.com}} and a direct drive equatorial mount\footnote{Astelco NTM-500, \url{www.astelco.com}}. Due to the large camera sensor, a field of view of $0.32^\circ \times 0.27^\circ$ is achieved while maintaining a resolution of about 0.5 arcseconds per pixel.

Publicly available TLE data\footnote{Two Line Elements, see e.g.\ \url{http://www.celestrak.com/NORAD/elements/}} or the more accurate CPF predictions\footnote{Consolidated prediction format, as used by the International Laser Ranging Service, see \url{http://ilrs.gsfc.nasa.gov/data_and_products/predictions/}} are used to track the target object. Even when using less accurate TLE data, the object can be captured reliably due to the large field of view. The object is identified in the camera image as the only point-like structure (stars appear as streaks). Its offset from the target point in the image is used to correct the pre-programmed trajectory of the mount. Using a PID control loop with appropriate settings, the object can usually be moved to the target point within \SI{30}{s}, and kept there with less than two arcsecond (RMS) deviation. This target point is usually near the image centre, and is previously adjusted to direct the returning laser photons onto the rather small single photon detector (see also section \ref{sec:receiver}). The tracking system is sensitive enough to see objects as small as standard CubeSats in Low Earth Orbit under favourable conditions (magnitude 12 to 13). Since the tracking system relies on the control loop for its accuracy, it is currently limited to track visible objects during dusk or dawn. More details on the tracking system can be found in \cite{Hampf2014}.

\subsection{Laser transmitter system}
\label{sec:transmitter}

\begin{figure}[tb]
   \centering
   \includegraphics[width=1\columnwidth]{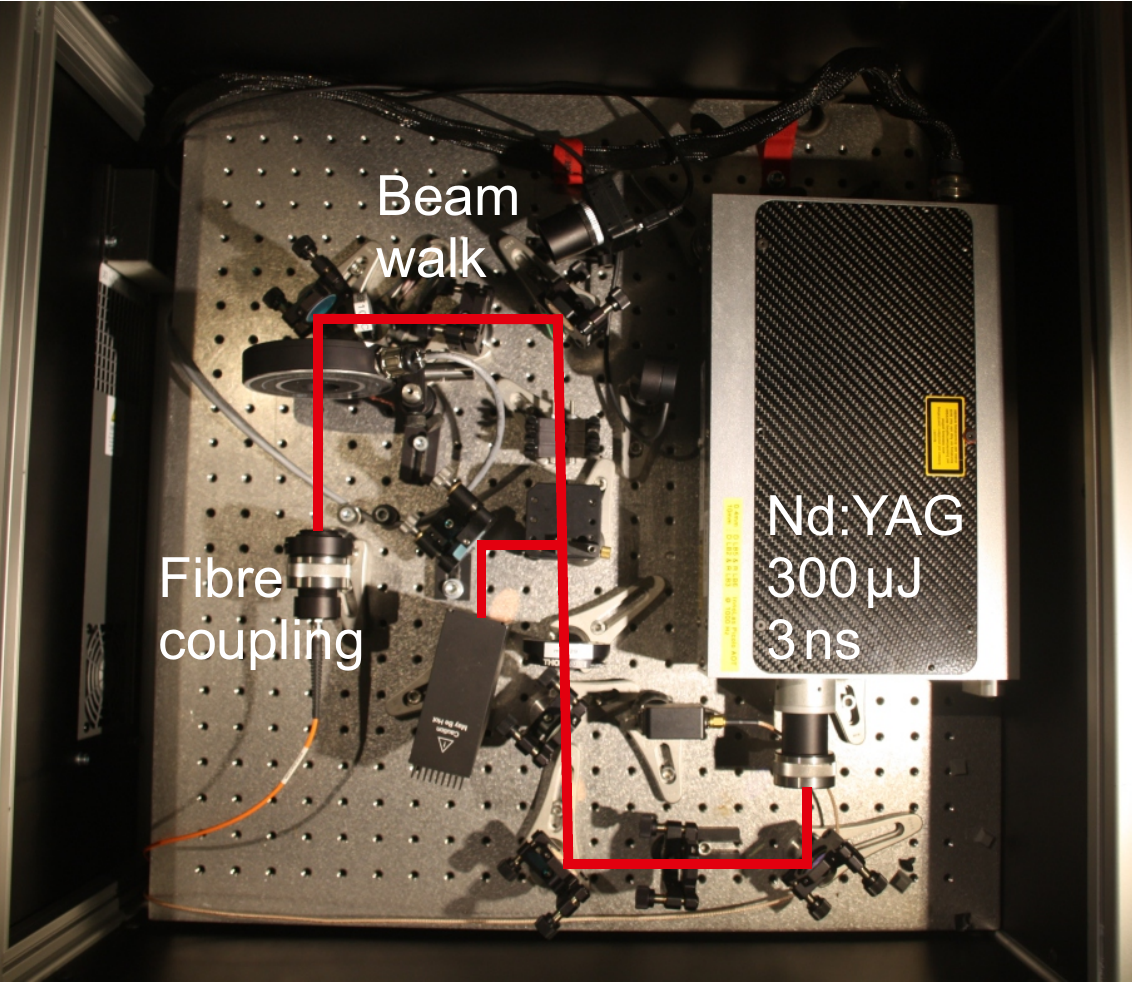}
   \caption{Picture of optical set-up for coupling the \si{\nano\second} laser pulses into the multi mode fibre.}
   \label{optical_setup_breadboard}
\end{figure}

The laser transmitter consists of three parts: the laser, the fibre with the corresponding coupling optics and the beam shaping unit mounted on the astronomical mount (see figure \ref{telescope_image}).

An active Q-switched Nd:YAG MOPA system\footnote{Innolas picolo AOT 1 MOPA, \url{http://www.innolas-laser.com/}} operated at \SI{1064}{\nano\meter} is used as laser pulse source. Linear polarized laser pulses with durations of \SI{3}{\nano\second} (FWHM) and energies up to \SI{300}{\micro\joule} are generated. The repetition rate can be varied between \SI{1}{\kilo\hertz} and \SI{10}{\kilo\hertz}. 

Figure \ref{optical_setup_breadboard} shows the optical set-up used for fibre coupling. To adjust the transmitted laser pulse energy the light passes a combination of half wave plate and polarizing beam splitter. In order to measure the pulse energy a beam splitter guides a small fraction of the light onto a power meter. A beam walk consisting of two tip/tilt adjustable mirrors is used for independent beam position/direction control. This is necessary to couple the light into the multi mode fibre (M14L05, Thorlabs). More information about the fibre as well as the fibre coupling can be found in section \ref{Fiber}. 

\begin{figure}[tb]
   \centering
   \includegraphics[width=1\columnwidth]{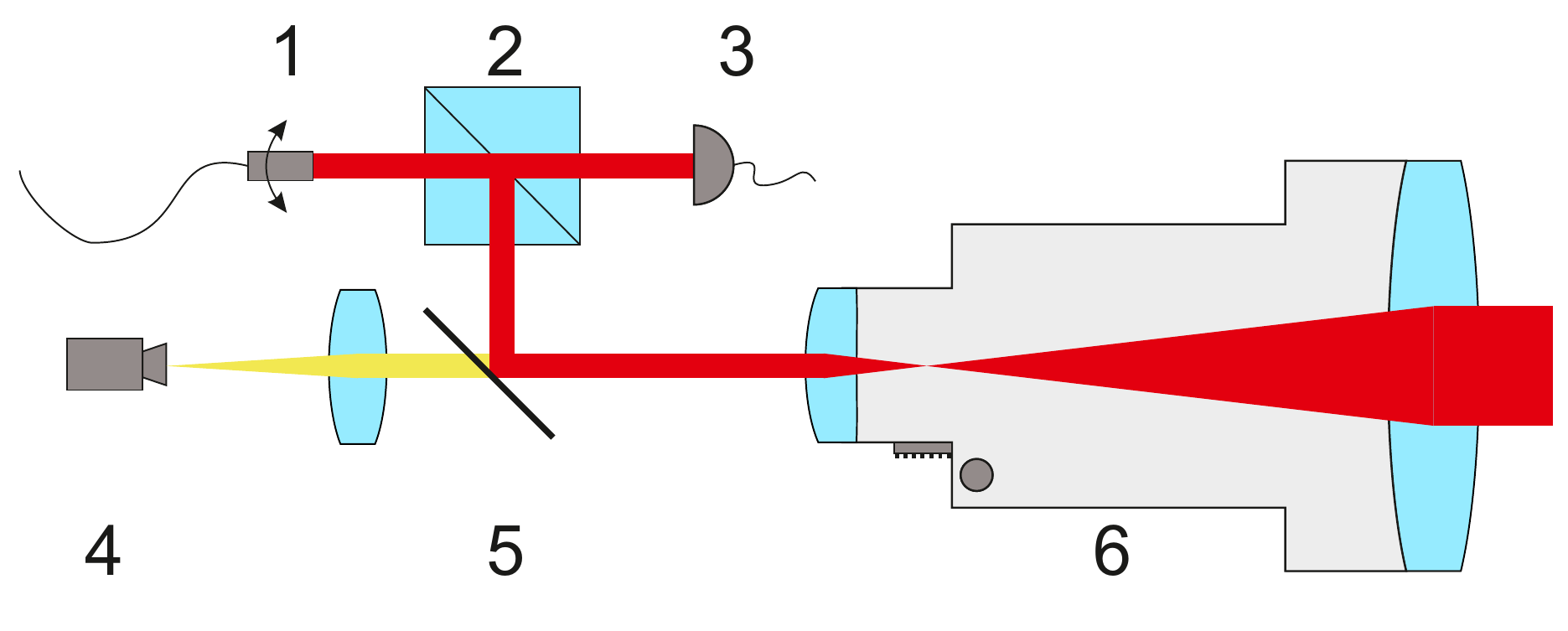}
   \caption{Optical set-up of the beam shaping optic. (1) fibre collimator mounted in beam steering unit, (2) 90:10 beam splitter, (3) energy monitoring, (4) transmitter camera, (5) dichroic mirror (R$>$\SI{95}{\percent} @ $\lambda>\SI{805}{\nano\meter}$), (6) Kepler telescope for beam expanding}
   \label{optical_setup_transmitter}
\end{figure}

After propagation through the fibre the light is collimated by a fibre collimator (F810SMA-1064, Thorlabs). A subsequent 90:10 (R:T) cube beam splitter guides a small portion of the light onto a power meter. This allows to continuously record the laser pulse energy transmitted into the atmosphere as well as the fibre transmission. The main fraction of the light is reflected by the beam splitter onto a dichroic mirror with a cut-off wavelength of \SI{805}{\nano\meter}. Wavelengths above this cut-off wavelength are reflected. Thus the collimated laser beam is redirected towards the expanding telescope as shown in figure \ref{optical_setup_transmitter}. 

This refractive Keplerian telescope consists of a plano-convex lens with a focal length of \SI{60}{\milli\meter} and a Fraunhofer achromat with a focal length of \SI{500}{\milli\meter} and a diameter of \SI{10}{\centi\meter}. As a result the beam is expanded to match the output aperture. The Fraunhofer achromat is part of a commercial telescope and mounted in a tube containing a manual focuser (see figure \ref{telescope_image}). With this focuser the distance between the plano-convex lens and the Fraunhofer achromat can be varied to control the divergence of the outgoing beam.

A beam steering unit is realized by placing the fibre collimator in a tip/tilt mount. To avoid beam position variations at the output during deflecting the beam, the fibre collimator is placed in the conjugate plane of the Fraunhofer achromat. With DC servo motors (Z806, Thorlabs) the beam can be deflected by more than \SI{1}{\milli\radian} with an accuracy of \SI{5}{\micro\radian} at the output of the laser transmitter. Thus optical misalignment between the tracking telescope and the beam shaping optic as well as small pointing errors caused for example by mechanical deformations are compensated in the sub Hz regime.

For alignment of the laser beam with respect to the optical axis of the tracking telescope a camera is used. The alignment procedure is described in \cite{Leif2015} in more detail.

\subsection{Fibre}
\label{Fiber}

It was already shown that even single mode fibres (SMF) are able to guide laser pulses of several ten micro joules with pulse durations of \SI{3}{\nano\second} \citep{Leif2015}. While SMFs show excellent beam quality \citep{KAN09} the pulse energies in the several ten micro joule regime are quite close to the fibre damage threshold \citep{SMI09}. Since our station is not air conditioned and vibration isolated long-term stable SMF coupling is hard to achieve. One possibility to overcome these restrictions is using MMFs with core diameters on the order of several ten micrometers. 

The used fibre has a core diameter of \SI{50}{\micro\meter} and a numerical aperture (NA) of 0.22. For an efficient fibre coupling the diameter of the focused beam must be smaller than the core diameter. Furthermore the maximum angle of incidence onto the fibre has to be smaller than its acceptance angle. 

A plano-convex lens with a focal length of \SI{50}{\milli\meter} fulfils both fibre coupling requirements. A diffraction limited focal diameter of \SI{30}{\micro\meter} and an NA of the focused beam of 0.023 is obtained.

While MMFs withstand higher laser peak powers than SMFs the achievable laser beam divergence is worse \citep{AND04}. With the fibre collimator used in the current set-up a divergence half angle of about \SI{1}{\milli\radian} was simulated using the commercial optic design software ZEMAX. This corresponds to a divergence half angle of \SI{125}{\micro\radian} at the output of the transmitter telescope. The simulation corresponds well to the measured divergence presented in section \ref{Divergence_estimation}.

\subsection{Receiver channel}

\label{sec:receiver}

\begin{figure}[tb]
   \centering
   \includegraphics[width=1\columnwidth]{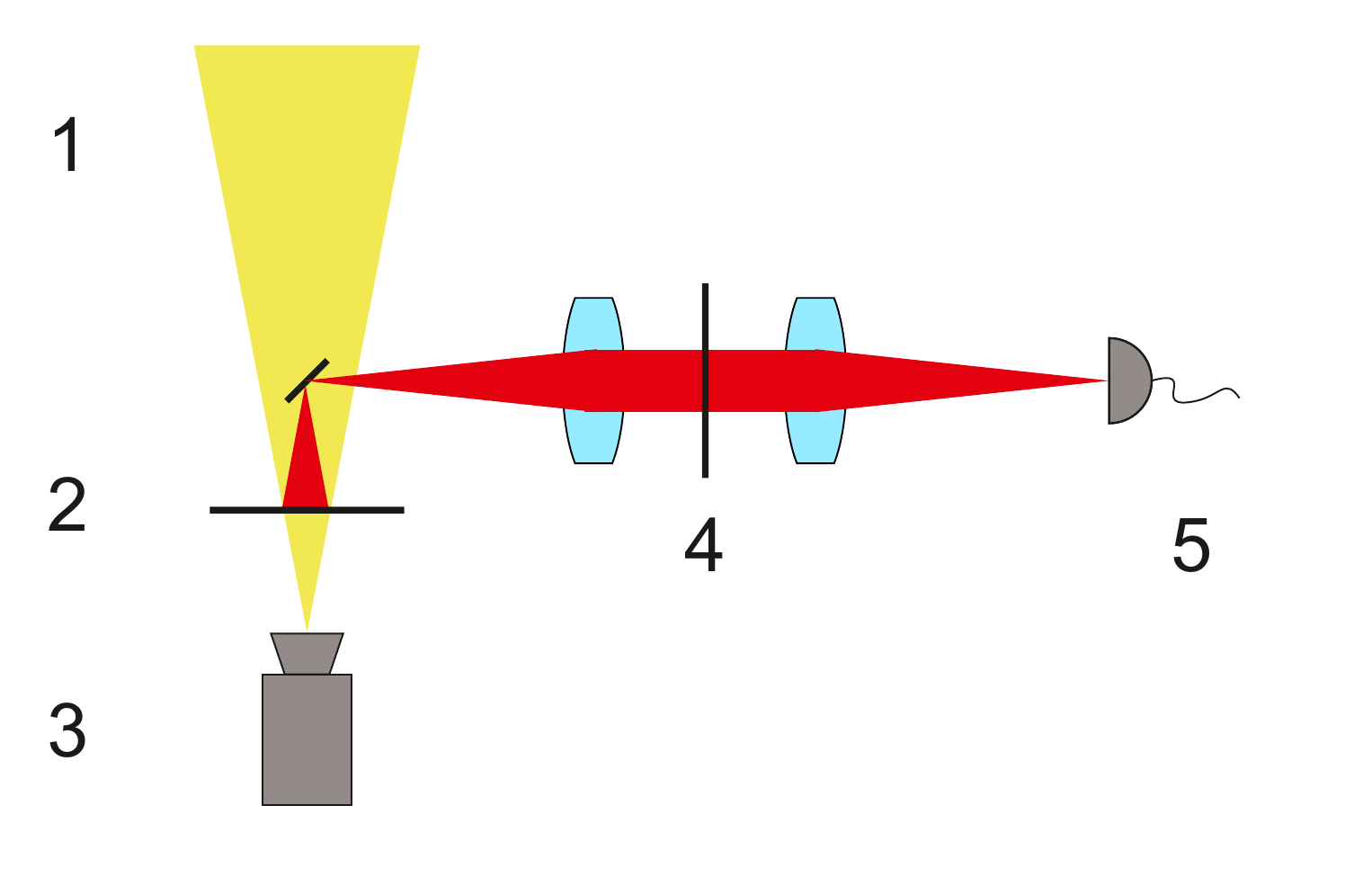}
   \caption{Optical set-up of the receiver unit: (1) incoming light from the mirror telescope, (2) dichroic mirror (R$>$90\% @ $\lambda > \SI{750}{\nano\metre}$), (3) tracking camera, (4) \SI{1064}{\nano\metre} bandpass filter, (5) single photon detector.}
   \label{receiver_beam_splitter}
\end{figure}

Laser photons returning from the target object are captured by the mirror telescope, which contains a dichroic beam splitter near its focal plane\footnote{While this beam splitter is custom built, commercial alternatives exist in on-axis guiding systems, e.g.\ \url{www.innovationsforesight.com/product-category/on-axis-guider}}. While all visible light passes through to the tracking camera, infrared photons are reflected onto a side port. There, a relay lens group guides the light onto the single photon detector. To reduce aberrations and to focus as much light as possible onto the detector, the second lens is of aspherical shape. Between the two lenses, a \SI{1064}{\nano\metre} bandpass filter (FWHM \SI{8}{\nano\metre}) is inserted to reduce noise (see figure \ref{receiver_beam_splitter}). 

An InGaAsP/InP SPAD\footnote{idQuantique id400, no longer available. \url{www.idquantique.com/}} (single photon avalanche photodiode) is used as single photon receiver. It features a detection efficiency of up to 30\% at \SI{1064}{\nano\metre}, while the dark count rate is well below \SI{2}{\kilo\hertz}. A circuit for active quenching, and a discriminator with TTL output are included in the detector module. The dead time is set to \SI{10}{\micro\second}, but can be reduced at the expense of a higher afterpulse rate. Due to its low noise rate, the detector can be operated without gating in our current set-up. However, gating is supported by the detector module, and can be used to reduce the amount of noise events especially for bright objects.

With its diameter of \SI{80}{\micro\metre} the detector covers a field of view of about six arcseconds. Therefore, the achieved two-arcsecond tracking accuracy is just enough to keep the object on the detector. The whole system of beam splitter, camera and detector is integrated into a rigid unit to avoid any movement of the detector against the camera, especially while slewing the telescope. To increase stability, the beam splitter and the relay optics do not contain any mechanical adjustment possibilities. To centre the object's image onto the detector area, the target point in the camera is adjusted accordingly. For the calibration, the telescope is directed onto a star, and the detector rate is maximised by moving the telescope slightly. The star position determined using this procedure is used as target point.

\subsection{Data acquisition (DAQ)}

The data acquisition system is used to record start and stop signals from the system and relate them to UTC. A PicoHarp 300 event timer\footnote{by PicoQuant, \url{http://www.picoquant.com/}} is used for the time of flight measurement. The start signal is generated by a photodiode on the transmitter breadboard, the stop signal by the id400 single photon detector in the receiver channel. TTL pulses for the laser trigger and the detector gating are generated by an FMC-DEL card\footnote{\url{http://www.ohwr.org/projects/fmc-delay-1ns-8cha/wiki}} in a small White Rabbit network \citep{Serrano:1743073}. 
In this network, a White Rabbit Switch coupled to a dedicated timing GPS module\footnote{Jackson Labs Fury GPS, \url{www.jackson-labs.com}} serves as grandmaster clock. It distributes the time to the FMC-DEL card, which in turn sends synchronisation signals to the PicoHarp event timer. Thus, all time tags are synchronised to UTC to better than \SI{100}{\nano\second}, while time differences are measured with an accuracy of better than \SI{1}{\nano\second}. More details, and a schematic of the set-up, can be found in \cite{Leif2015}.

The data acquisition system is designed to avoid the need for any real-time programming. Instead, all time critical calculations are performed on hardware level, and high level information is sent to the PC via USB (PicoHarp) and PCI bus (FMC-DEL). Therefore, the control software can run on any current PC with a standard Linux distribution. The system is also designed for high data rates: Each individual component can work with data rates of up to \SI{1}{\mega\hertz}, while system performance is currently limited by the software to about \SI{10}{kHz}. Improvements on the software are underway with the goal to handle system data rates of more than \SI{100}{kHz}.

\section{Measurements and results}
\label{results}

\subsection{Divergence estimation}
\label{Divergence_estimation}

\begin{figure}[tb]
   \centering
   \includegraphics[width=1\columnwidth, height=200px]{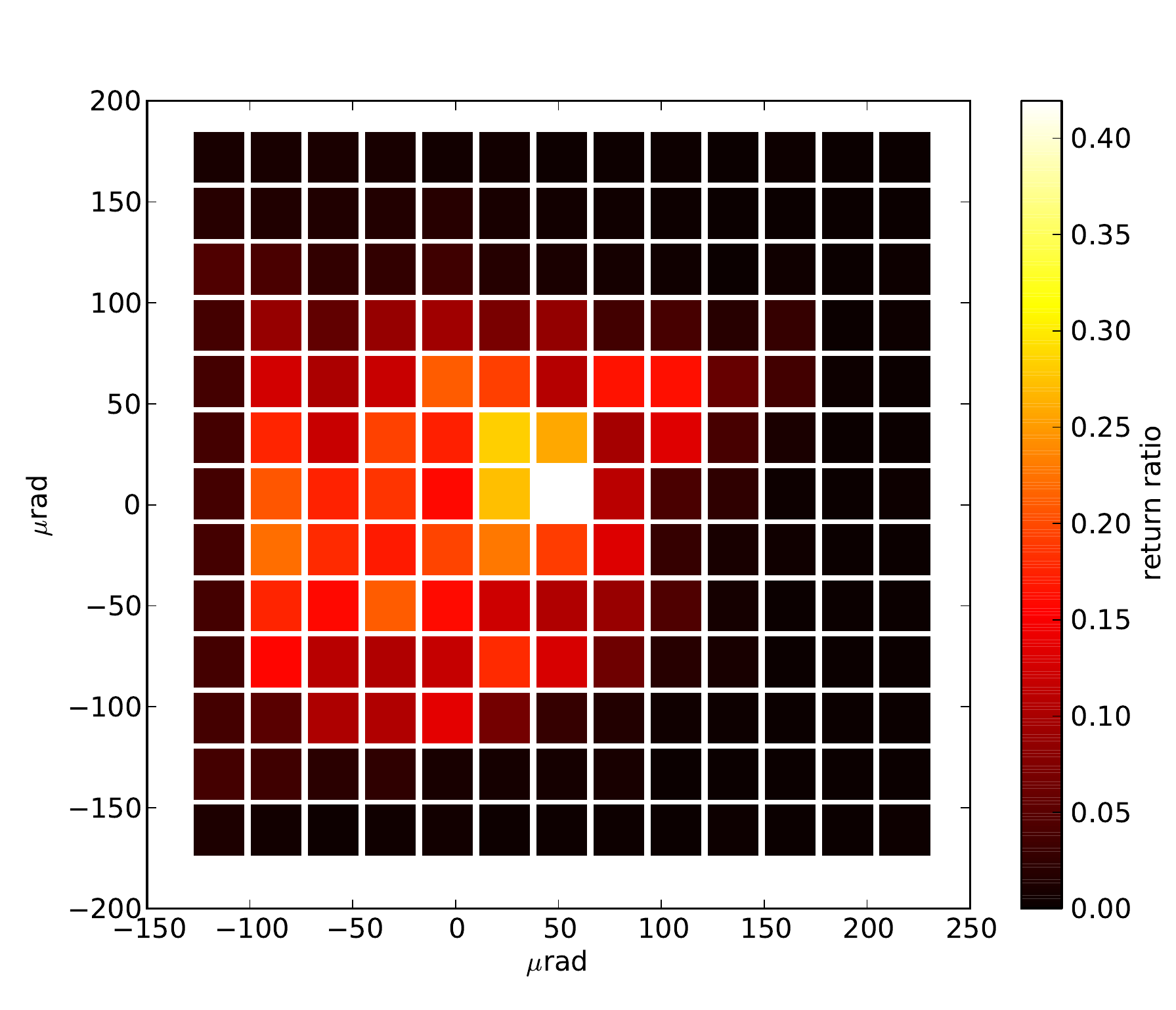}
   \caption{Beam profile measured using the beam steering unit and the retroreflector at 5.3 km distance. The colour scale gives the measured return ratio at the respective direction (number of detected return pulses / number of emitted laser pulses, \SI{1}{\second} average).}
   \label{fig:divergence}
\end{figure}

Two local test targets have been installed for calibration and evaluation of the system performance. Within the dome, a mirror and a diffuse screen can be used to guide transmitted laser pulses into the receiver. This set-up is used mainly to determine cable and system delay times. 

The second target consists of a \SI{12.7}{\milli\metre} retroreflector mounted on a hilltop at about \SI{5.3}{\kilo\metre} distance. It is used to test all system components (laser source, transmitter, receiver, DAQ and software) in combination. Furthermore, the divergence of the laser beam, which is a critical quantity in the set-up, can be estimated using this target. For this, the receiver is pointed exactly at the reflector, while the laser beam is moved slightly using the beam steering unit. The maximum return ratio is adjusted to well below one photon per pulse by decreasing the laser power accordingly. Then, while moving the (large) beam across the (small) reflector, the beam profile can be sampled by measuring the return ratio. While this measurement is subject to many uncertainties (e.g.\ by air turbulences, the non-perfect movement of the beam steering motors, a varying detector efficiency etc.) it delivers a reasonable idea of the beam divergence.

A measurement conducted in this way is shown in figure \ref{fig:divergence}. The laser was operated at \SI{3}{\kilo\hertz} repetition rate, and the output pulse energy was attenuated to about \SI{3}{\pico\joule} to obtain mostly single photon returns. Fitting a normal distribution to the measured values yields a divergence estimate of about \SI{140}{\micro\radian} half angle. 
While this is sufficiently small for measurements on rather near targets, as shown in section \ref{Ranging_to_satellites}, a still smaller divergence is needed for targets above LEO. Since the used fibre with a rather large core diameter and NA contributes significantly to the divergence, it will be replaced in the future by a more suitable one (see also section \ref{fibre_based_SLR}).

\subsection{Ranging to satellites}
\label{Ranging_to_satellites}

\begin{figure}[tb]
   \centering
   \includegraphics[width=1\columnwidth, height=200px]{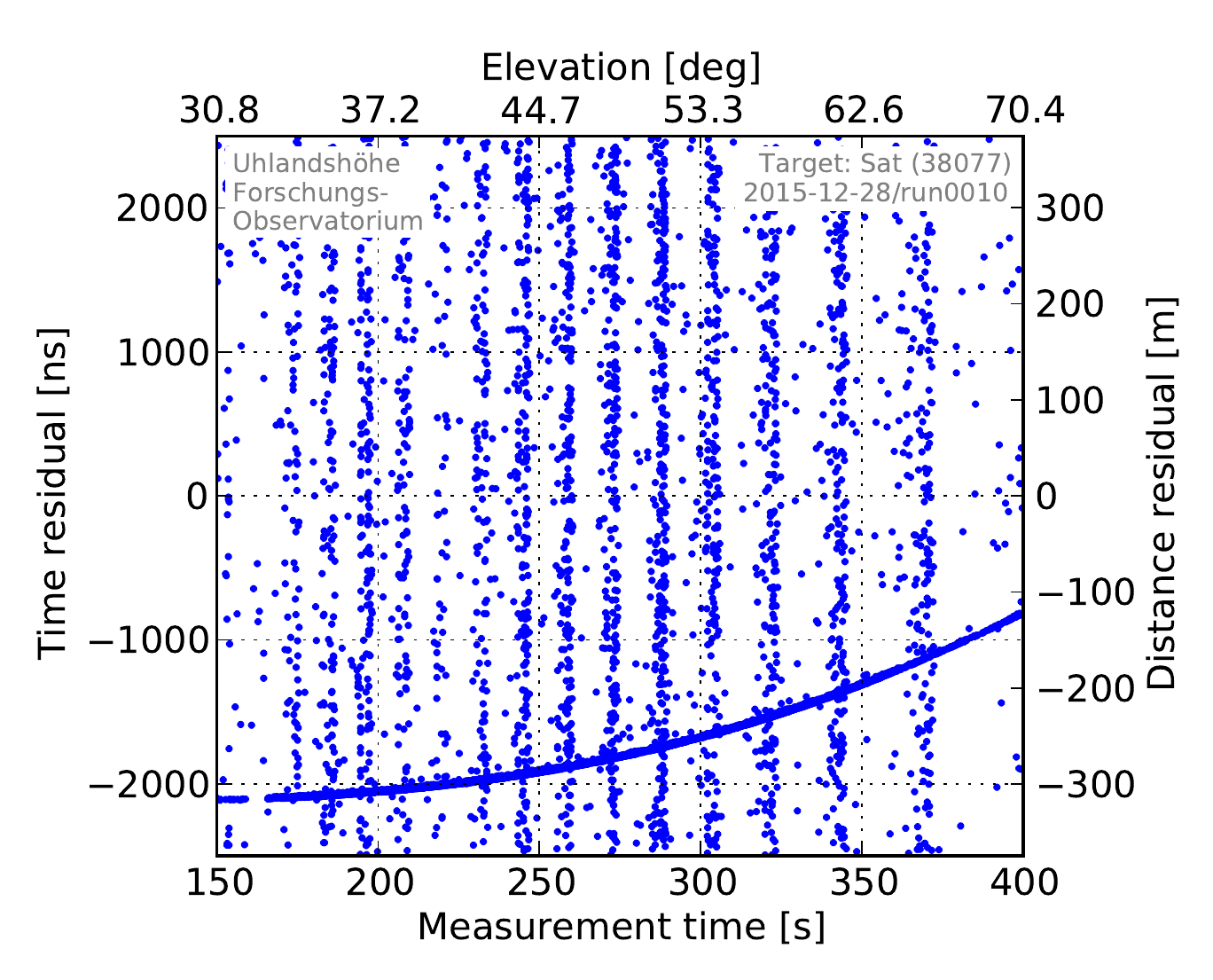}
   \caption{Time-of-flight residuals measured from a pass of satellite LARES (Norad ID 38077) on 2015-12-28, with $t_0=$17:45:31 UTC. The pass covers elevations from 30 to 70 degrees, and ranges from 2300 to 1550 kilometres. Residuals are given against TLE prediction.}
   \label{ranging_lares}
\end{figure}

First successful SLR measurements with this set-up have been carried out in December 2015. The laser was operated at a repetition rate of \SI{3}{\kilo\hertz} and an average power of about \SI{75}{\milli\watt} at the output aperture, i.e.\ at a pulse energy of \SI{25}{\micro\joule}. 

Figure \ref{ranging_lares} shows the ranging plot of a pass of satellite LARES, with a return ratio of about 3\% to 6\% (return rate of roughly 100 to \SI{200}{\hertz}). Due to the high repetition rate and the low noise of the detector, the signal is clearly visible despite the low return ratio. In this run, both laser and detector have been operated free-running, i.e.\ without external triggering / gating by the FMC-DEL card. Atmospheric backscatter is clearly visible, and becomes especially pronounced at times at which the expected time of flight (ToF) is a multiple of the pulse interval. It is planned to reduce the amount of backscatter by using a more flexible laser triggering scheme with pulse collision avoidance in the future.

During the first weeks of operation, several other cooperative LEO targets have been observed, with distances between 500 and \SI{2500}{\kilo\metre} and return rates between 50 and \SI{1.5}{\kilo\hertz}.
\section{Towards fibre based high performance SLR}
\label{fibre_based_SLR}

In the previous section successful ToF measurements to LEO objects are reported using a MMF to guide nanosecond laser pulses of few \SI{10}{\micro\joule} pulse energy from the laser source to a Keplerian telescope. The single shot precision of this set-up is mainly determined by the laser pulse duration to about \SI{0.5}{\meter}. Another problem is the rather high laser beam divergence of \SI{140}{\micro\radian} half angle. While this is sufficient for LEO laser ranging, the system can not compete with the performance of most stations in the international laser ranging service (ILRS).

The laser ranging stations in the ILRS network measure routinely to objects up to the medium Earth orbit (MEO) with cm single shot accuracies. By averaging return events so called normal points (NP) are calculated \citep{PEA}. Thereby distance precisions of the order of a few millimetres are achieved. The following paragraphs discuss potential ways to reach the ILRS performance with a fibre based SLR station.

One limitation of the presented system is the maximum working range of only about \SI{2500}{\kilo\meter}. While it has been shown that pulse energies of about \SI{30}{\micro\joule} are sufficient to range up to MEO \citep{KIR15}, the beam must be well collimated for these long ranges. In the present system, a rather large divergence is caused by the used MMF with high NA and large fibre core diameter. First of all a lot of modes are excited in such a fibre \citep{SAL91}. This leads to a spatial partially coherent laser beam at the fibre output. Such a laser beam shows a higher diffraction limited divergence than the Gaussian beam emitted from a SMF \citep{TAK85,FRI82}. Furthermore the fibre collimator introduces optical aberrations leading to a non perfect collimated laser beam at the output. The influence of these aberrations increase with increasing fibre NA and core diameter. 

Both effects can be significantly reduced by using MMFs with smaller NA and core diameter. Another approach is using single mode fibres enabling nearly diffraction limited Gaussian beams. While conventional SMFs show best divergence performance, the pulse energy is limited due to the small core diameter in the range of \SI{5}{\micro\meter}. Even for the used nanosecond laser pulses the peak intensities are close to the damage threshold of the material. This pulse energy restriction can be relaxed by using innovative SMF designs like large mode area fibres \citep{KNI98}, hollow core photonic crystal fibres \citep{HUM04,SHE04}, kagome fibres \citep{DUM14} or active fibres \citep{KOP00,LIM02}. Thus ranging to satellites in MEO seems possible with MMFs as well as with innovative SMFs.

Beside the divergence issue the achievable NP precision is important for high performance SLR. This precision can be estimated by the error of the arithmetic mean which is given as the single shot precision divided by the square root of conducted measurements. Thus the NP precision can either be improved by increasing the single shot precision or by increasing the number of measurements \citep{DEG03}. The latter one can be achieved by increasing the laser repetition rate which can be realized with the presented system. Even the current repetition rate of \SI{3}{\kilo\hertz} enables NP precisions significantly below the single shot precision. Possibilities to increase the repetition rate further will be investigated to reach the intrinsic precision limit of the system.

To improve the single shot precision, shorter laser pulse durations can be used. Today, most ILRS stations perform ranging with picosecond laser pulses. The challenge of this approach is the higher peak intensity of picosecond pulses compared to nanosecond pulses of the same pulse energy. Again, multi kHz ranging is beneficial since it allows for lower pulse energies. Modal dispersion of picosecond pulses in MMFs does not present a problem. Whereas this effect spreads the pulse by about \SI{130}{\pico\second} in case of a step index MMF the pulse spread is only \SI{1}{\pico\second} in case of a gradient index MMF (fibre length of \SI{5}{\meter}) \citep{SAL91}.  Thus picosecond laser ranging seems possible with an MMF or perhaps innovative SMF based laser transmitter.

Another drawback of the current system is the need to visually acquire the object, which limits the operation time to dusk and dawn. This issue is however solely due to the relatively poor blind tracking accuracy of the used mount. It is not related to the fibre-based approach and can be overcome using a more accurate mount. Since no coud\'{e} path is needed, a large selection of suitable mounts is readily available.

In summary, it seems feasible to achieve high performance and high accuracy SLR with the proposed design.

\section{Conclusion and outlook}
\label{sec:conclusion}

The experiment has shown that it is possible to conduct satellite laser ranging using a modular set-up and rather inexpensive commercial hardware. In several aspects, it has been demonstrated how limitations in the hardware can be compensated by innovative system design and software: To mitigate the pulse energy restriction posed by the optical fibre, high repetition rates are used. To handle the poor blind pointing of the mount, an automated closed loop tracking has been developed. To achieve optical alignment between receiver and transmitter telescope, the laser beam direction is constantly adjusted by the beam steering unit. We believe that this approach offers a lot of potential for SLR and related fields.

Following the proof of concept presented here, the system will be improved with two main goals: First, system sensitivity should be improved to obtain clear signals also from more distant objects, including navigation satellites. Among other measures, this may be achieved by a reduction of the beam divergence, the improvement of the beam pointing accuracy, or an increase in the pulse repetition rate or pulse energy. Second, the accuracy and precision of the system shall be evaluated and verified, with the ultimate goal of approaching current ILRS standards.

\section{Acknowledgements}
The authors would like to thank the volunteer members working at the historical observatory in Stuttgart (Schw\"abische Sternwarte e.V.) for their hospitality and cooperation.

\bibliographystyle{model5-names}\biboptions{authoryear}
\bibliography{references}
\end{document}